\documentclass[12pt]{article}
\usepackage{amsmath,amsfonts}
\usepackage{algorithmic}
\usepackage{algorithm}
\usepackage{array}
\usepackage[caption=false,font=normalsize,labelfont=sf,textfont=sf]{subfig}

\usepackage{cite}
\usepackage{pgfplots}
\pgfplotsset{compat=1.18} 
\usepackage[inkscapeformat=png]{svg}
\usepackage{newtxtext,newtxmath}
\usepackage{lipsum}     
\usepackage{placeins}   
\usepackage{needspace} 
\usepackage{graphicx}

\usepackage[letterpaper,margin=1in]{geometry}

\usepackage{xcolor}
\renewenvironment{abstract}
	{\quotation}
	{\endquotation}

\date{}

\makeatletter
\renewcommand{\fnum@figure}{\textbf{Figure \thefigure}}
\renewcommand{\fnum@table}{\textbf{Table \thetable}}
\makeatother

\usepackage{url}

\def\scititle{Robust NbN on Si-SiGe hybrid superconducting–semiconducting microwave quantum circuit}
\title{{\bfseries\boldmath \fontsize{16}{12}\selectfont \scititle} }

 \author { \fontsize{12}{8}\selectfont{
  Paniz Foshat,
  Samane Kalhor,
  Shima Poorgholam-khanjari,
  Douglas Paul},\\
   \fontsize{12}{8}\selectfont{
  Martin Weides,
   and Kaveh Delfanazari$^{\ast}$}\\
  \small \textit{James Watt School of Engineering, University of Glasgow, Glasgow, UK}\\
  \small  $^\ast$Corresponding author: \textcolor{blue} {Kaveh.Delfanazari@glasgow.ac.uk}  \: Dated: 30/09/2025
}
\begin{document} 

\maketitle

\begin{abstract} \bfseries \boldmath
Advancing large-scale quantum computing requires superconducting circuits that combine long coherence times with compatibility with semiconductor technology. We investigate niobium nitride (NbN) coplanar waveguide resonators integrated with Si/SiGe quantum wells, creating a hybrid platform designed for CMOS-compatible quantum hardware. Using temperature-dependent microwave spectroscopy in the single-photon regime, we examine resonance frequency and quality factor variations to probe the underlying loss mechanisms. Our analysis identifies the roles of two-level systems, quasiparticles, and scattering processes, and connects these losses to wafer properties and fabrication methods. The devices demonstrate reproducible performance and stable operation maintained for over two years, highlighting their robustness. These results provide design guidelines for developing low-loss, CMOS-compatible superconducting circuits and support progress toward resilient, scalable architectures for quantum information processing.
\end{abstract}
\noindent
\label{Back Ground}
\\
Silicon-germanium (Si$_x$Ge$_{1-x}$), a semiconductor alloy of silicon (Si) and germanium (Ge), emerges as a promising candidate for various types of qubit implementations in quantum information processing \cite{sammak2019shallow,scappucci2021germanium}. Its compatibility with current CMOS technology and promising electronic properties \cite{goley2014germanium} make it a versatile platform for various types of hybrid qubits \cite{strickland2024characterizing,luthi2018evolution,aguado2020perspective} and cryogenic electronics devices \cite{delfanazari2024large,delfanazari2024quantized,delfanazari2018chip,delfanazari2017chip}, such as quantum dot qubits \cite{lawrie2020quantum}, spin qubits \cite{petersson2012circuit}, gatemon qubits \cite{zheng2024coherent} and Josephson junction transistor \cite{vigneau2019germanium}. 
SiGe heterostructures can be engineered to form quantum dots, allowing for precise control over qubit performance and scalability \cite{brauns2016highly,pitsun2020cross}. In spin qubits, SiGe's long coherence times \cite{kobayashi2021engineering,wang2022ultrafast} and low nuclear spin density \cite{moutanabbir2024nuclear} enable efficient manipulation of electron spin states for quantum operations. Consistently,  
a recent study investigated the coherence times of conventional transmon qubits using niobium superconductors on Si/Si$_{0.85}$Ge$_{0.15}$/Si substrate stack layers \cite{sandberg2021investigating}. These qubits exhibited coherence times exceeding 100 $\mu$s, demonstrating that this heterostructure supports coherence times comparable to conventional qubits fabricated on high-resistivity silicon. \par
In parallel, SiGe-based Josephson junction transistors have gained significant attention for their CMOS compatibility, positioning them as promising candidates for scalable quantum device architectures \cite{vigneau2019germanium,aggarwal2021enhancement,valentini2024parity,tosato2023hard}. 
Initial research indicated that two-dimensional hole gas (2DHG) quantum wells based on Ge confinement (Ge/SiGe) could potentially be used to fabricate voltage-tunable transmon qubits and Josephson junction transistors \cite{hinderling2024direct,zhuo2023hole,sagi2024gate,vigneau2019germanium}. However, the application of Ge quantum wells in transmon qubits has been restricted by losses due to inherent defects within the strain layers of SiGe quantum well wafers, thickened SiGe barrier, crystalline quality of the Ge, and high concentrations of Ge \cite{nigro2024demonstration}.
For instance, charge noise, often originating from two-level fluctuators (TLF) within the semiconductor, can potentially diminish quantum processors' coherence and dephasing time \cite{paquelet2023reducing, kopas2021low}. The sources of these TLF come from impurities found in various locations: within the quantum well itself, the semiconductor barrier, and the semiconductor superconductor interface \cite{connors2020charge,culcer2009dephasing,dekker1991spontaneous}. Additionally, achieving high Ge concentration becomes complicated due to a large lattice mismatch to the Si substrate \cite{nigro2024demonstration}.
As a result, recent research efforts have shifted towards investigating (two-dimensional electron gas) 2DEG quantum wells where Si quantum wells are confined between SiGe stack layers (Si/SiGe) \cite{huang2024understanding,paquelet2023reducing,degli2024low}. This type of wafer presents advantages, including a lower Ge concentration and a thinner SiGe barrier, which may result in reduced microwave loss and higher coherence times when used in voltage-tunable transmon qubits. Despite the potential advantages, this area remains underexplored. \par Throughout this paper, we refer to 2DEG SiGe wafers as Si/SiGe, and 2DHG SiGe wafers as Ge/SiGe.
This work presents the design, fabrication, and RF measurement of NbN coplanar waveguide resonators on n-doped Si/Si$_{0.7}$Ge$_{0.3}$ wafers to assess their potential for the microwave readout of qubit states in semiconductor, superconductor or potentially topological-based quantum devices \cite{serra2020evidence,microsoft2025interferometric}. The first section presents a detailed discussion of the novel NbN CPW resonator fabrication process on the Si/SiGe substrate. We introduce a novel fabrication process that reduces the interaction of the CPW resonators' electric field with the charge noise in the semiconductor quantum well. Following this, we present atomic-scale images of the NbN-sputtered wafer used in this study. Subsequently, we report measurement results from the fabricated chip, obtained in a dilution refrigerator (DR) under varying power and temperature conditions. 
The measured data exhibit an internal quality factor ($Q_i$) of approximately 1100, consistently maintained across repeated cooldowns over two years, which leads to achieving long-term stability and robustness, and highlights the potential of this wafer platform for semiconductor-based qubits. Finally, we compare our chip with recent CPW resonators fabricated using Ge/SiGe heterostructures to benchmark device characteristics and assess the relative advantages of Si/SiGe-based wafers. Through this research, we provide insights into the dominant loss mechanisms in Si/SiGe wafers and explore their potential to be used in voltage-tunable, CMOS-compatible qubits.
\begin{figure}[!t]
  \centering
  \includegraphics[width= 6 in]{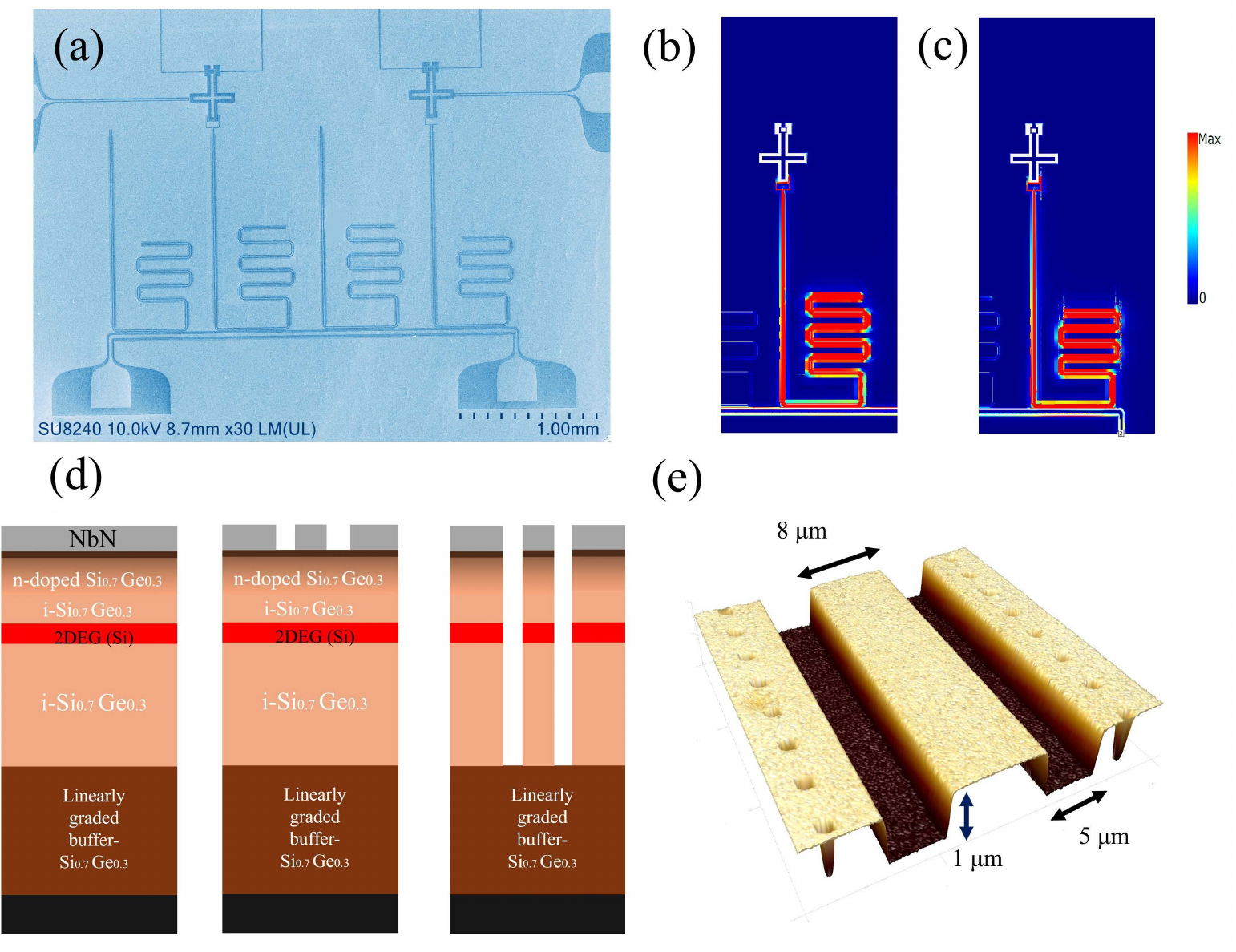}
  \caption{ (a) False colour SEM image of fabricated SiGe chip. (b,c) Surface current density magnitude $J_{s}$ (A/m) for the first and second resonant modes, respectively, obtained from Sonnet simulations. (d) NbN-Si/SiGe CPW resonator step-by-step fabrication process, from left to right, includes: NbN sputtering, superconducting film etch, and SiGe buffer layer etch. (e) The AFM line profile of NbN CPW resonator fabricated on the Si/SiGe wafer.}
  \label{fig:figsamp}
\end{figure}

\section*{Hybrid NbN/Si-SiGe Microwave Quantum Circuit Fabrication}

Our superconducting microwave circuits consist of microwave resonators capacitively coupled to a common coplanar waveguide fabricated on a 100 nm NbN film sputtered on a Si/SiGe wafer. The device layout consists of four CPW resonators coupled to a common feedline. Two $\lambda$/4 CPW resonators were designed with the open end capacitively coupled to a large capacitor (for qubit design purposes) and the short end strongly coupled to the feedline. These capacitors are part of the intended qubit–resonator architecture; however, in this work, we restrict the discussion to these two resonators' characteristics, while the qubit-level analysis will be reported separately. Figure~\ref{fig:figsamp}(a) presents a scanning electron microscopy (SEM) image of the device following fabrication.
In addition, Fig.~\ref{fig:figsamp}(b) and (c) show false-colour maps of the surface current density magnitude $J_{s}$ (A/m) for the two $\lambda$/4 resonators at their fundamental frequencies, obtained from Sonnet simulations. The currents are strongly confined to the resonators and feedline, with minimal leakage into the surrounding ground plane.
Two additional $\lambda$/2 resonators are also present in the chip design. In Sonnet simulations without considering the contribution of kinetic inductance, their resonance frequencies lie outside the 4–8 GHz band. In fabricated devices, finite kinetic inductance and film losses would further shift their response. As these resonators are not relevant to the present study, their data are not reported here; a detailed analysis will be presented elsewhere.

Fabrication of these CPW resonators began with a buffered oxide etchant (BOE) dip to remove the native oxide from the surface of the Si/Si$_{0.7}$Ge$_{0.3}$ substrate. Immediately after removing the contamination layer, the chip was placed in the MP 600 S Plassys sputter tool's load lock to prepare for NbN superconducting film deposition. After the load lock was pumped down, the sample was transferred to the main chamber to sputter 100 nm NbN. A 100 nm-thick NbN film was deposited on a Si/Si$_{0.7}$Ge$_{0.3}$ wafer via reactive DC sputtering of a Nb target in the main chamber. The process was carried out for 15 minutes using 25 sccm of Ar and 3.0 sccm of N$_2$, with a base current of 0.85 A.
After removing the wafer from the load lock of the sputtering machine, the ZEP resist was spun and baked at 180 $^o$C on a hotplate for four minutes. Then, the CPW resonators were patterned on NbN superconducting film with standard e-beam lithography, followed by CF$_4$/Ar anisotropic dry etch.\par 
In the next step, a novel deep etching process was subsequently applied to the CPW gap regions, etching through the heterostructure stack and stopping at the linearly graded relaxed Si$_{0.7}$Ge$_{0.3}$ buffer layer. The slow deep etching proceeded with SF$_{6}$ and C$_{4}$F$_{8}$, leading to 1 $\mu$m etched depths in the gaps of NbN CPW resonators. This additional substrate etch step sets apart our process from conventional CPW resonators fabrication \cite{foshat2025characterizing}, where only the superconducting layer is patterned and etched. While similar approaches have previously been explored for devices on Si substrates \cite{bruno2015reducing}, applying this novel fabrication step to a  Si/Si$_{0.7}$Ge$_{0.3}$ wafer offers a way to reduce intrinsic loss. By etching into the substrate in the CPW resonators' gap regions, the substrate-vacuum interfaces and quantum well charge noise sources are shifted away from areas of high electric field in CPW resonators, potentially improving coherence in SiGe-based microwave superconducting circuits. 
Figure~\ref{fig:figsamp}(d) illustrates the step-by-step fabrication process of the CPW resonators, as described above, progressing from left to right. The visual sequence includes NbN sputtering, patterning, and etching of the superconducting film and finally etching of the underlying buffer layers.
Figure~\ref{fig:figsamp} (e) indeed presents the Atomic Force Microscopy (AFM) profile of the fabricated CPW resonator chips, implying the deep etching of the CPW resonators' gap.  \par
We selected one of these chips to capture images with the Transmission Electron Microscopy (TEM) tool, allowing us to showcase the crystalline properties of the wafer heterostructures and assess the quality of NbN sputtering. We obtained a cross-sectional lamella sample by employing the Focus Ion Beam (FIB) tool after depositing platinum (Pt) on the chip to protect it from structural damage. The cross-sectional images from this lamella were captured using Scanning Transmission Electron Microscopy (STEM) techniques at various magnifications. \par
\begin{figure}[!t]
  \centering
  \includegraphics[width= 6 in]{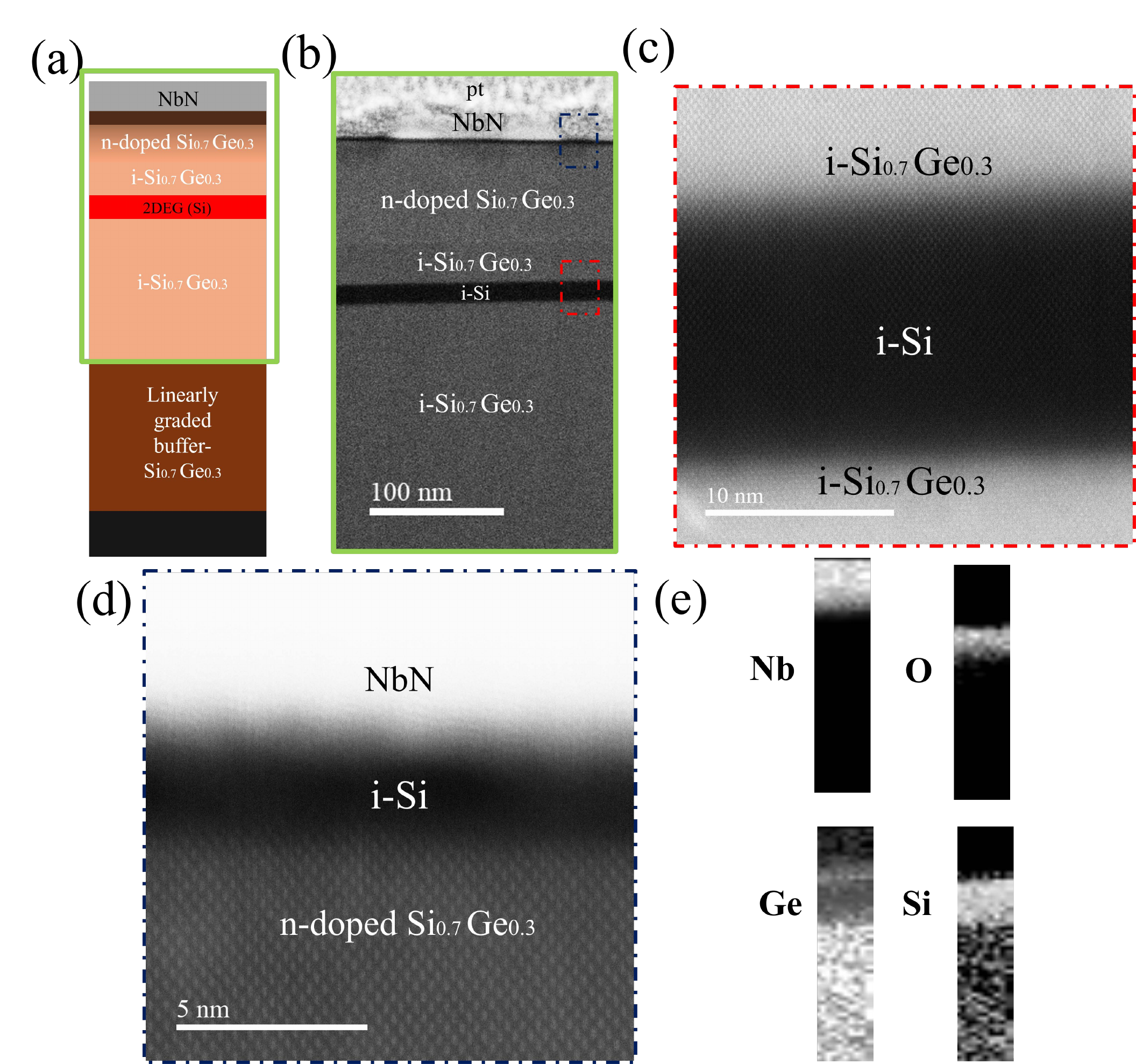}
  \caption{(a) Schematic of the SiGe wafer with sputtered NbN top layer. (b) Low-resolution cross-sectional STEM images of the wafer. (c) HAADF images of the Si constraint quantum well layer between the Si/SiGe buffer layer. (d) High-resolution HAADF images between n-doped layer Si$_{0.7}$Ge$_{0.3}$, the Si cap layer, and the sputtered NbN layer. The scale bar is 5 nm. (e) Elemental map of this area, including Ge, Nb, O, and Si.}
  \label{fig:fig4_3}
\end{figure}

Figure~\ref{fig:fig4_3} (a) presents the schematic of a Si/Si$_{0.7}$Ge$_{0.3}$ wafer where the heterostructures were grown on a Si (001) substrate with the CVD method, followed by NbN sputtering on the top. From bottom to top of the wafer, illustrated in Fig.~\ref{fig:fig4_3} (a), is comprised of a thick SiGe relaxed buffer layer made of graded Si$_{1-x}$Ge$_{x}$ followed by compressive strained Si$_{0.7}$Ge$_{0.3}$ layer with constant Ge concentration. A strained two-dimensional Si layer, around 7 nm, has been deposited on the Si$_{0.7}$Ge$_{0.3}$ compressive strained layer to form the active region of the quantum well. On top of the Si quantum well, a second Si$_{0.7}$Ge$_{0.3}$ strained layer was grown. The High Angle Annular Dark Field (HAADF) image at Fig.~\ref{fig:fig4_3} (c) displays perfect epitaxial growth of Si$_{0.7}$Ge$_{0.3}$-Si-Si$_{0.7}$Ge$_{0.3}$ layers, with atomically resolved layers and lattice-matched growth.
To improve the transport properties and mobility, a 50 nm n-type Si$_{0.7}$Ge$_{0.3}$ was formed on top of Si$_{0.7}$Ge$_{0.3}$ top strained layer, resulting in a high electron mobility ($\approx$ 30000 cm$^2$/Vs at 1.5 K). Finally, a 4 nm Si cap layer was deposited on top of the wafer to prevent damage and contamination to the SiGe layers.    
Figure~\ref{fig:fig4_3} (d) presents a HAADF image of the wafer’s top region, viewed from the bottom, including the n-doped Si$_{0.7}$Ge$_{0.3}$ constraint layer, Si cap layer and NbN. Figure~\ref{fig:fig4_3} (d) illustrates perfect crystallinity in SiGe layers; however, the crystallinity is lost in the Si cap layer. In Fig.~\ref{fig:fig4_3} (e), the Electron Energy Loss Spectroscopy (EELS) image of this area demonstrates the evidence of oxygenation in the cap layer, likely caused by exposure to air during wafer transport to the sputtering tool. Figure~\ref{fig:fig4_3} (e) depicts the elemental maps of these Ge, Si, and NbN elements. \par
Subsequently, after dicing our wafer, we took optical images of each device before glueing and mounting them into sample boxes.
Two samples, named sample 1 (S1) and sample 2 (S2), were found to be free of visible scratches, short circuits, or unintended connections. Both samples were designed with the same geometrical parameters, including the resonator's length, width (8 $\mu$m), gap (5 $\mu$m) size and a uniform 100 nm thick NbN superconducting film.\par We want to note that despite identical design parameters, measured resonance frequencies exhibited minor variations between samples, likely due to fabrication-related inconsistencies.
After preparing the sample boxes, we mounted them in the dilution refrigerator and conducted four separate cooldown cycles over a two-year period. The first two cooldowns were performed one month apart, followed by a third measurement eight months later. The final cooldown was carried out two years after the initial run to assess the long-term stability and reliability of these circuits. The first, second, third and fourth cooldowns are denoted by the acronyms $1^{st}$, $2^{sc}$, $3^{rd}$ and $4^{th}$, respectively, in the following sections.

\subsection*{Cryogenic spectroscopy of hybrid NbN/Si-SiGe superconducting circuits}
Understanding the influence of power and temperature on CPW resonators is fundamental to designing and fabricating hybrid quantum circuits with high coherence.
We found fundamental frequencies at $f_{r1,s1}$ $\simeq$ 5.04 GHz, $f_{r2,s1}$ $\simeq$ 5.57 GHz, $f_{r1,s2}$ $\simeq$ 5.08 GHz and $f_{r2,s2}$ $\simeq$ 5.60 GHz, all measured at a base temperature of 50 mK. We employed a notch-type circle fitting model, as described in Ref.~\cite{probst2015efficient}, to extract resonator parameters, such as $Q_i$, coupling quality factor ($Q_c$), and loaded quality factor ($Q_l$) from $S_{21}$ transmission data.\par
\begin{figure}[t]
 \hspace{-1cm}\includegraphics[]{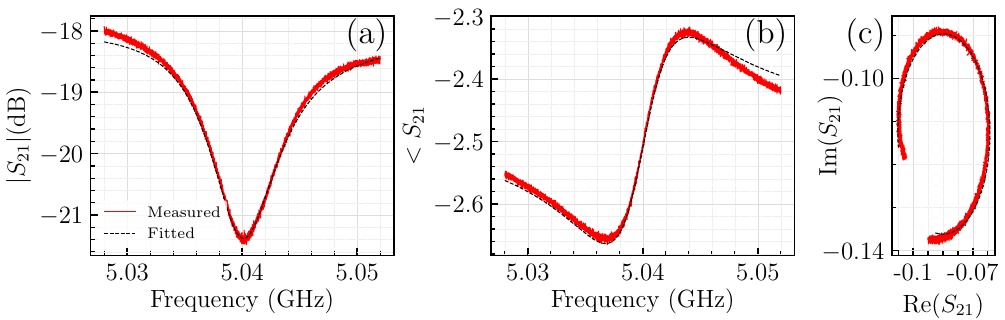}
 \vspace{0.1 cm}
 \hspace{-0.9 cm}\includegraphics[]{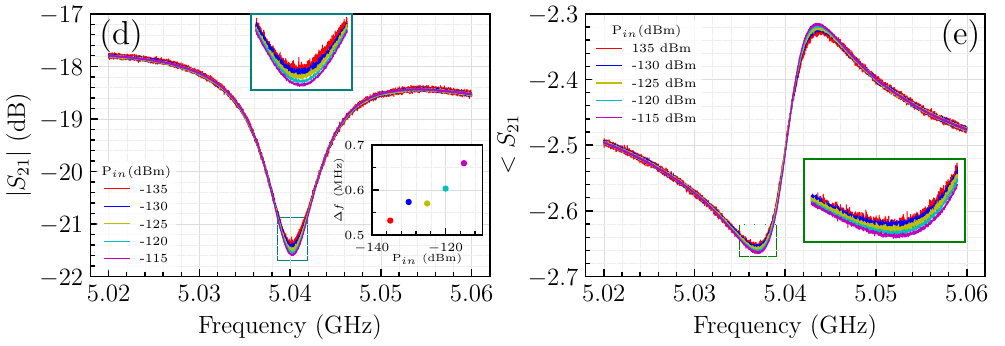}
 \caption{ Measured (red) and fitted (black) data of notch-type circuit model for (a) $|$S$_{21}|$ (dB), (b) $\langle$ S$_{21}$, and (c) parametric plot at P$_{in}$ = -115 dBm. (d) $|$S$_{21}|$ (dB) and  (e) $\langle$ S$_{21}$ at $f_{r1,s1}$ = 5.04 GHz, with the power range -140 dBm $<$ P$_{in}$ $<$ -115 dBm at $T$ = 50 mK. All data are extracted from S1 at $3^{rd}$ measurements. }
 \label{fig:fig4_6}
 \end{figure}

Figure~\ref{fig:fig4_6} (a) and Fig.~\ref{fig:fig4_6} (b) depict the measured and fitted $|$S$_{21}|$ and $\langle$ S$_{21}$ of the CPW resonator around $f_{r1,s1}$, measured at $T$ = 50 mK using notch type circuit model. Similarly, Fig.~\ref{fig:fig4_6} (c) shows the complex $S_{21}$ data for S1 at $T$ = 50 mK,  illustrating the close agreement between the measured data and the circle fit. The strong alignment between fitted and experimental data confirms the accuracy and reliability of the fitting procedure.\par
Building on this, we carried out power spectroscopy measurements across the power range of -140 dBm $<$ P$_{in}$ $<$ -115 dBm for both chips. In this context, P$_{in}$ = P$_{VNA}$ + P$_{att}$, where P$_{VNA}$ is the VNA output power, and P$_{att}$ presents attenuation at room temperature and within the dilution refrigerator. Figure~\ref{fig:fig4_6} (d) and Fig.~\ref{fig:fig4_6} (e) highlight the changes in $|$S$_{21}|$ and $\langle$ S$_{21}$ across the applied power range. As evident in Fig.~\ref{fig:fig4_6}(d), increasing the power results in a more prominent resonance dip. A closer view of this plot further reveals a gradual upward shift in the resonance frequency, with a total shift of up to 700 kHz observed across the measured power range. This behaviour reflects the impact of two-level system (TLS) defects on the resonator's response, where increased input power leads to TLS saturation and a corresponding upward shift in the resonance frequency.

 \begin{figure}[t]
 \centering
 \hspace{-1cm} \includegraphics{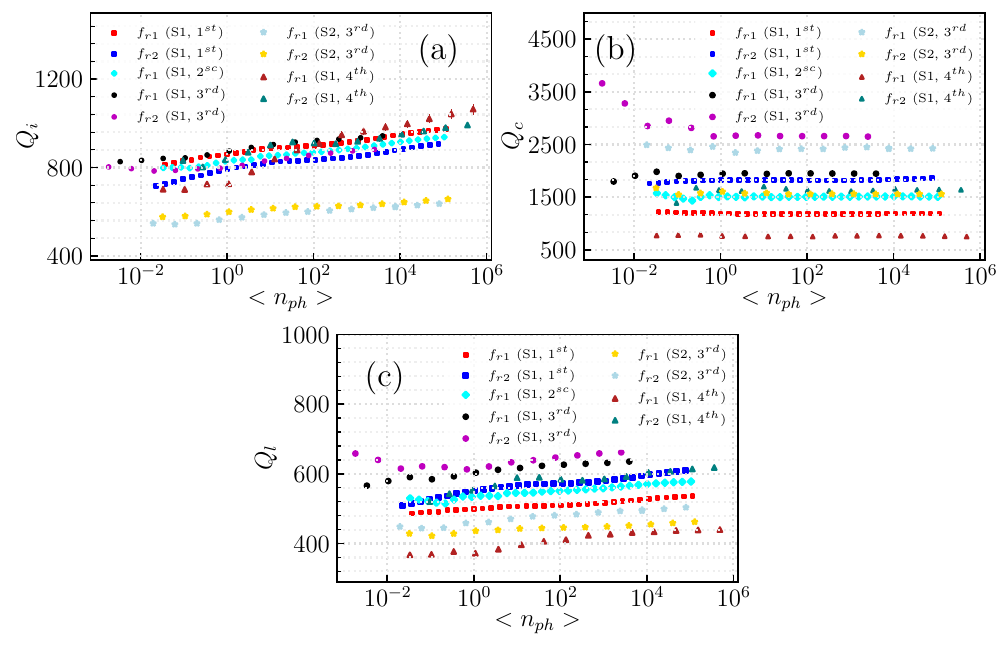}
 \caption{ Measured $Q_i$ (a), $Q_c$ (b) and $Q_l$ (c) at the fundamental frequency for both chips in the four different cooldowns.}
 \label{fig:fig4_9}
 \end{figure}
\subsection*{Power-dependent Microwave Spectroscopy of CPW Resonators}

In this section, we present power-dependent microwave spectroscopy measurements of the CPW resonators, performed during the fourth independent cooldowns.\par Figure~\ref{fig:fig4_9} (a,b,c) illustrate the relation between $Q_i$, $Q_c$ and $Q_l$ to the photon number ($< n_{ph} >$), respectively. These measurements correspond to the same cooldown cycles discussed previously. Figure~\ref{fig:fig4_9} (a) shows that the $Q_i$ of sample S1 increases from 860 in the single photon regime to 1064 at higher powers, demonstrating a relatively high $Q_i$ compared to similar devices based on semiconductor wafers.\par The absence of $Q_i$ saturation at low photon numbers in the first two measurements was unexpected, especially given the clear saturation observed in NbN/Si CPW resonators \cite{foshat2025characterizing}. To further investigate potential TLS-related loss mechanisms, we conducted a third measurement with reduced input power, which visibly showed that $Q_i$ saturates in the ultra-low power regime. It is also intriguing that $Q_i$ remained stable in the third measurement (performed eight months later) and the fourth (conducted two years later), indicating no measurable degradation over time. This long-term consistency highlights the robustness and ageing resistance of our chip design, particularly against surface oxide regrowth \cite{verjauw2021investigation}. In the same figure, the $Q_i$ of S2 is measured to be in the range from approximately 500 in the single photon regime to 600 at higher powers.\par
Figure~\ref{fig:fig4_9} (b,c) depict the $Q_c$ and $Q_l$ measurement results of this hybrid circuit, respectively. These data confirm that $Q_c$ remains largely constant across the entire power range and depends on the design as expected.\subsection*{RLC circuit model}
Here, we aim to discuss the corresponding RLC circuit model for this new generation of CPW resonators, beginning with an approximation of the dielectric constant of the Si/SiGe wafer. The dielectric constant for the Si$_{1-x}$Ge$_x$ layers is approximated using the expression $\epsilon_{SiGe}$ = $\epsilon_{Si}$ + 4.5$x$. Given that the dielectric constant of Si, $\epsilon_{Si}$, is 11.7 and the mole fraction of Ge is 0.3, $\epsilon_{SiGe}$ is approximately 13.05 \cite{sandberg2021investigating,levinshtein2001properties}. Therefore, the permittivity of the wafer is:
\begin{equation}
    \epsilon_{sub} = \frac{d_{Si} \epsilon_{Si} +  d_{SiGe} \epsilon_{SiGe}} {d_{Si}+d_{SiGe}}
\end{equation}
Where $d_{Si} = 378 \: \mu$m is Si substrate thickness, and $d_{SiGe} = 2 \: \mu$m is Si$_{0.7}$Ge$_{0.3}$ layer thickness. For a quarter wavelength resonator, the resonance frequency has an inverse relation to its length, where $f_r =  c / \sqrt{\epsilon_{eff}} \times 1/4L$. Here, $c$ is the speed of light, $\epsilon_{eff}$ is the effective permittivity, and $L$ is the length of the CPW resonators. 
With a complex conformal mapping equations $\epsilon_{eff}$ would be defined as \cite{zhang2024high}:
\begin{equation}
    \epsilon_{eff} =  \frac{1+\epsilon_{sub} \tilde{K} }{1+\tilde{K}}
\end{equation}
\begin{equation}
   \tilde{K} = \frac{K(k^{\prime})K(k_3)}{K(k)K(k_3^{\prime})}
\end{equation}
Here, $K$ is the complete elliptical integral of the first kind, with $k$=$w/(w+2g)$ when $w$ is the width of the resonator, and $g$ is the CPW resonator's gap. In addition, $k_3 = tanh(w\pi/4d)/tanh((w+2g)\pi/4d)$, $k^{\prime}=\sqrt{1-k^2}$ and $k_3^{\prime} = \sqrt{1-k_3^2}$. In this formula, $d$ is the final thickness of wafer, meaning, $d= d_{Si}+d_{SiGe}$\cite{zhang2024high}.
\begin{table}[b]
\caption{Table lists the resonator lengths, designed resonance frequencies, measured resonance frequencies, and $Q_i$ values in the single photon regime for both chips.}
\begin{center}
\scalebox{0.8}{
\begin{tabular}{||c|c|c|c|c|c|c|c|c||} 
 \hline
 \textbf{Sample}&\textbf{length} & \textbf{$f_{r,\:design}$} & \textbf{$f_{r,\:measure}$} & $L_k$& $L_g$&$C_g$&\textbf{$Q_{i}$} & R \\ [0.5ex] 
 \hline\hline
 S1 &4643 $\mu$m & 6.40 GHz  & 5.57 GHz& 0.142$\mu$H/m & 0.44 $\mu$H/m& 0.159 nF/m & $813 $ & 16.08 $\Omega$/m \\
 \hline
 S1&5084 $\mu$m & 5.85 GHz & 5.04 GHz & 0.152 $\mu$H/m& 0.44 $\mu$H/m& 0.159 nF/m & $776$ & 15.52$\Omega$/m \\ [0.5ex] 
 \hline
 S2 &4643 $\mu$m & 6.40 GHz  & 5.60 GHz& 0.135 $\mu$H/m & 0.44 $\mu$H/m& 0.159 nF/m & $ \approx 568$ & 22.8 $\Omega$/m \\
 \hline
 S2 &5084 $\mu$m & 5.85 GHz & 5.08 GHz & 0.143 $\mu$H/m& 0.44 $\mu$H/m& 0.159 nF/m & $599$& 19.95 $\Omega$/m\\ [0.5ex] 
 \hline
\end{tabular}}
\end{center}
  \label{tab:table_permitivity2}
\end{table}

From a geometrical perspective of CPW resonators, geometrical inductance ($L_g$) and geometrical capacitance ($C_g$) can be determined as follows \cite{goppl2008coplanar}:
\begin{equation}
 L_g =  \frac{\mu_0}{4} \frac{K(k')}{K(k)}
\end{equation}

\begin{equation}
 C_g =  4 \epsilon_0 \epsilon_{eff} \frac{K(k)}{K(k')} 
\end{equation}
With the above equations, $\epsilon_{eff}$ is 6.35 and Table~\ref{tab:table_permitivity2} shows kinetic inductance ($L_k$), geometrical inductance, and capacitance based on our measurement data for both chips. To extract resistance ($R$) in these CPW resonators, we used the equation $R = \frac{Z_{eff}}{Q_i L}$ \cite{wallraff2004strong} and $Z_{eff} = \sqrt{(L_{k}+L_{g})/C_{g}} $. The calculated RLC circuit model and its schematic for each resonance frequency are shown in Table~\ref{tab:table_permitivity2}.
\par
 
To evaluate our measurement results, we compared our measurement data in Table~\ref{tab:table4_2} with the most recent fabricated CPW resonators based on Ge/SiGe wafer, focusing on top buffer layer thickness, $Q_i$, $Q_l$ and $f_r$. Recent research demonstrates CPW resonators' behaviours during power sweep in an undoped Ge/SiGe wafer employing the forward-graded growth method \cite{valentini2024parity} and an undoped Ge/SiGe quantum well wafer using the reverse-graded growth method \cite{nigro2024demonstration}, respectively. Our study diverts attention to the CPW resonators based on Si/SiGe wafers. Our results demonstrate a comparable $Q_i$
with lower measurement error bars compared to CPW resonators fabricated on reverse-graded undoped Ge/SiGe wafers. The lower error bar is required for fast and accurate readout of a semiconductor-based qubit. Additionally, the extracted $Q_i$ is similar to the CPW resonators fabricated on a forward-graded Ge/SiGe wafer with a thick SiGe top buffer layer, while being lower than the $Q_i$ obtained from CPW resonators fabricated on a thin SiGe top buffer layer. The higher $Q_i$ in the wafer with thin SiGe top layers may be attributed to the proximity effect inducing superconductivity in the quantum well, thereby suppressing the scattering effect and potentially increasing $Q_i$. 
\begin{table}[h]
 \caption{Microwave superconducting circuits based on SiGe quantum well.}
\begin{center}
\scalebox{0.8}{
 \begin{tabular}{||c c c c c c c||} 
 \hline
 \textbf{Article} & \textbf{Substrate} & \textbf{Spacer distant} & \textbf{$f_r$} & \textbf{$Q_l$} & \textbf{$Q_{i \space < n_{ph} >\space \approx \space 1 }$}  & \textbf{$Q_{i \space < n_{ph} > \space \approx \space 100 }$} \par  \\ [2ex] 
 \hline\hline

 Valentini \cite{valentini2024parity} & Undoped Ge/SiGe & 5 nm  & 5.97 (GHz) & $630 \pm 10$ & $2900 \pm  150$ & $4000 \pm 20$ \\ [2ex] 
 \hline
 Valentini \cite{valentini2024parity} & Undoped Ge/SiGe & 60 nm  & 6.14 (GHz) & $730  \pm  10$ & $1930  \pm  40$ & $2200  \pm 50$\\ [2ex] 
 \hline
 Nigro\cite{nigro2024demonstration} & Undoped Ge/SiGe & 55 nm  &  5-8 (GHz) & 550 & $900 \pm 100$ & $900 \pm 100$ \\[2ex] 
 \hline
Foshat (2025) & n-doped Si/SiGe & 77 nm  &  5.04 (GHz) & $488 \pm  10$ & $865 \pm  4.5$ & $899 \pm 5$\\[2ex]
\hline
Foshat (2025) & n-doped Si/SiGe & 77 nm  &  5.57 (GHz) & $552 \pm  25$ & $822 \pm 2.67$ & $851  \pm  2.75$\\[2ex]
\hline
\end{tabular}}
\end{center}
  \label{tab:table4_2}
\end{table}
We emphasise that, unlike previous studies that used aluminium, our work leverages NbN, a robust type II superconductor, to fabricate CPW resonators on Si/SiGe substrates. This choice enhances the devices' tolerance to magnetic fields, temperature and environmental fluctuations. Moreover, NbN’s superconducting gap frequency ($\approx$1 THz) empowers device operation in the THz regime, leading to a reduced footprint and the miniaturisation of hybrid quantum circuits.
\subsection*{Loss mechanism}
The overall loss in CPW resonators can be assessed by examining the specific contributions of different loss mechanisms, including dielectric, scattering, quasiparticle, and radiative losses \cite{foshat2025quasiparticle,foshat2025characterizing,poorgholam2025engineering}. 
 \begin{table}[!b]
 \caption{Fitted constant values of the TLS loss model to measured data.}
\begin{center}
 \begin{tabular}{||c c c c c||} 
 \hline
 \textbf{Sample} & \textbf{$\beta$} & \textbf{$n_c$} & \textbf{$Q_{TLS}^0$} & $Q_0$ \\ [0.5ex] 
 \hline\hline
 S1-$f_{r1} = 5.04$ & $0.16 \pm \:  0.06$  & $0.32 \pm \: 0.2$ & $5703 \pm 1025$ & $946 \pm 9.4$\\
 \hline
 S1-$f_{r2} = 5.57$ &  $0.3 \pm 0.15$ & $2.1 \pm 1.48$ & $6789 \pm \: 1099$ & $894 \pm \:  18$ \\ [0.5ex] 
 \hline
 S2-$f_{r1} = 5.085$ & $0.0791 \pm 0.0308$& $0.0104 \pm 0.0104$ & $2941 \pm 669$ & $693.8 \pm 32.43$\\
 \hline
 S2-$f_{r2} = 5.604$ & $0.0820 \pm 0.024$  & $0.115 \pm 0.055$ & $2342 \pm 330$ & $694 \pm 22$ \\ [0.5ex] 
 \hline
\end{tabular}
\end{center}
  \label{tab:table4_3}
\end{table}
In this section, we characterise NbN/SiGe CPW resonators' loss as a combination of TLS loss, quasiparticle loss and scattering loss as below:
\begin{equation}
\frac{1}{Q_i} = \frac{1}{Q_0} +\frac{1}{Q_{TLS}(T,P)} + \frac{1}{Q_{qp}(T)}
\end{equation}
The initial term represents the loss mechanism, which remains unaffected by variations in temperature and power, such as those coming from lattice mismatch, grain boundaries in SiGe layers and scattering centres in the quantum well. The second term is TLS loss, which is associated with TLS-related defects localised at the interface between the superconductor and substrate, as well as within the substrate and oxidation. The third term, quasiparticle loss, is negligible in our measurements at $T$ = 50 mK because the thermal energy at this temperature is much lower than the superconducting energy gap of NbN, preventing Cooper pair breaking. We modelled the $Q_i$ extracted from our chip using the TLS loss equation below to analyse its contribution to the overall loss. 
\begin{equation}
\frac{1}{Q_{TLS}} = \frac{1}{Q_{TLS}^0} \frac{tanh{(\frac{h f_r}{2 k_B T}})}{(1+\frac{< n_{ph} >}{n_{c}})^\beta}
\end{equation}
Here, $n_c$ is the critical photon number, $k_B$ is boltzmann constant, $\frac{1}{Q_{TLS}^0} = \delta_{TLS}^0$ is intrinsic TLS loss, and $\beta$ is a fitting value. \par
Table~\ref{tab:table4_3} presents the TLS loss equation model values that fit the measured data in Fig.~\ref{fig:fig4_9}, for both samples. This table demonstrates that $Q_{TLS}^0$ is higher than $Q_0$, verifying that while TLS losses contribute to dissipation, the dominant loss mechanism is more closely related to scattering centres and internal loss is not strongly dependent on photon number.

Many of these scattering loss centres originate from structural imperfections, such as grain boundaries, dislocations, and residual strain, which exist both at the superconductor film and within the wafer. However, unlike TLS losses, which are power- and temperature-dependent \cite{foshat2025characterizing,poorgholam2025engineering}, scattering losses persist constantly, even at millikelvin temperatures. Moreover, the low surface resistivity, R$_\square$ = 127 $\Omega/\square$ of the Si/SiGe wafer, introduces high conductivity compared to conventional high resistivity Si, leading to energy dissipation in the wafer.  
To support this, fabricating CPW resonators on Si/SiGe wafers with systematically varied growth conditions, doping levels, carrier densities ($n_s$), and surface uniformity would provide a controlled platform for extracting the contributions of each mechanism \cite{laroche2015scattering, madhavi2000low}.\par Recent theoretical work highlights the dependence of scattering loss on carrier density ($n_s$), connecting it to impurity concentration in the capping layer and surface roughness in Si/SiGe wafers \cite{huang2024understanding}.
This research can be achieved using microwave spectroscopy measurements of superconducting CPW resonators fabricated on Si/SiGe wafers with different $n_s$ during optimising strains, doping levels, surface roughness, lattice mismatch, charge carrier and percolation densities. For instance, in this study, we characterised the surface roughness and threading dislocation density of our chip and compared it with CPW resonators fabricated on a high-resistivity silicon substrate. This characterisation helps us understand how surface roughness affects the CPW resonators' properties. Further details are provided in the Appendix.
\subsection*{Temperature-dependent Microwave Spectroscopy of CPW Resonators}
Quasiparticles resulting from breaking Cooper pairs also contribute to the $Q_i$ value at higher temperatures, approximately at $T \gg T_c/10$.  
\begin{figure}[t]
 \centering
 \hspace{-1 cm}
  \includegraphics[width=6.5 in]{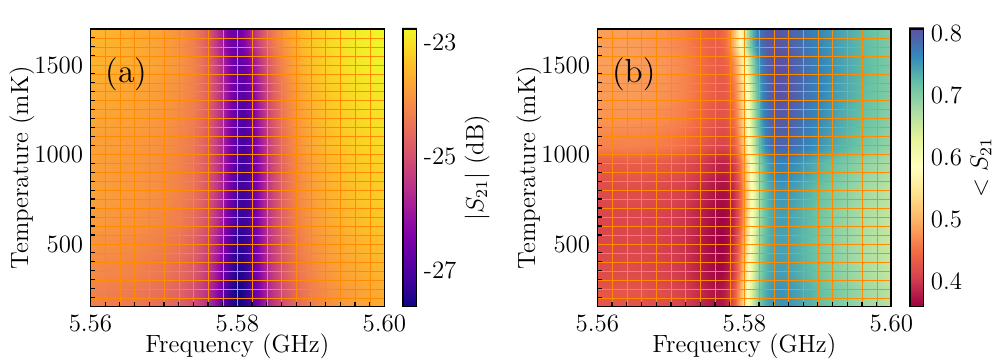}
  \caption{Measured frequency spectrum (a) and phase (b) for the temperature range between $T$ = 150 mK to 1700 mK with 100 mK step at $P_{in}$ = -115 dBm. Data are extracted from S1 at $2^{nd}$ measurements.}
  \label{fig:fig4_10}
\end{figure}


\begin{figure}[!b]
 \centering
  \hspace{-1 cm}\includegraphics[]{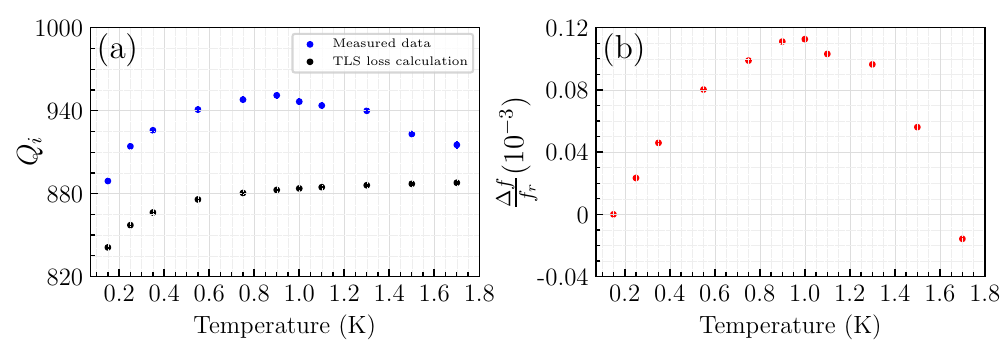}
  \caption{(a) $Q_i$ versus temperature, (b) $\Delta f$ of $f_{r2,s1}$ versus temperature at $P_{in}$ = -115 dBm. Both datasets are extracted from S1 at $2^{nd}$ measurements.}
  \label{fig:fig4_10_3}
\end{figure}
 In Fig.~\ref{fig:fig4_10} (a), the dips in $|$S$_{21}|$ at 1.7 K are shallower than at 150 mK, indicating the influence of quasiparticle losses. Additionally, the resonance frequency experiences a blue shift as the temperature increases from 150 mK to 900 mK, possibly due to TLS losses. Subsequently, from $T$ = 900 mK to 1.7 K, the resonance frequency undergoes a redshift, likely attributed to quasiparticle losses in hybrid circuits. This red shift occurs due to the increasing value of quasiparticle density, which alters the kinetic inductance of the superconducting film. For further analysis, we also calculated TLS loss in this temperature range in Fig.~\ref{fig:fig4_10_3} (a), with the constant values in Table~\ref{tab:table4_3}. This figure illustrates the variation between the extracted $Q_i$ and the calculated $Q_{TLS}$, indicating the presence of an additional temperature-dependent loss mechanism beyond TLS loss. Following that, we conducted power spectroscopy measurements at different temperature points ($T$ = 50 mK, 100 mK, 200 mK, 300 mK and 400 mK) to discern the underlying causes of temperature-dependent loss channels with a primary focus on TLS contributions and the potential role of scattering mechanisms.
\par
\begin{figure}[!t]
 \centering
 \hspace{-0.5 cm}\includegraphics[]{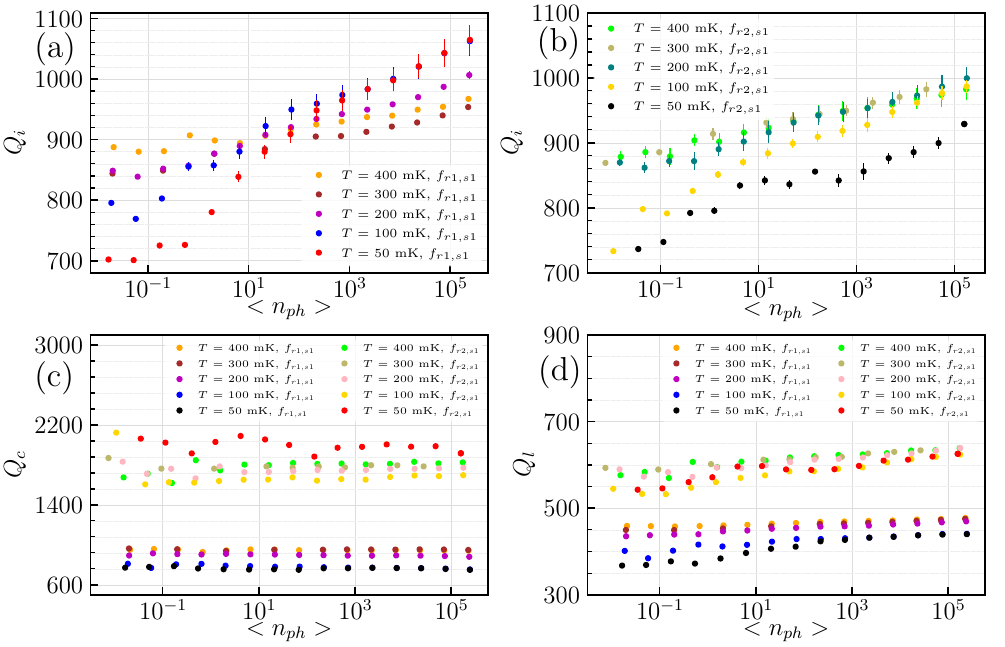}
 \caption{Extracted $Q_i$ as a function of photon number inside the resonator at (a) $f_{r1,s1}$ =  5.03 GHz and (b) $f_{r2,s1}$ = 5.56 GHz (c) $Q_c$ (d) $Q_l$ at -140 dBm  $\le$ $P_{in}$ $\le$ -70 dBm at $T$ = 50 mK, 100 mK, 200 mK, 300 mK and 400 mK. All data are extracted from S1 at $4^{th}$ measurements.}
 \label{fig:fig4_10_10}
 \end{figure}

 Figures~\ref{fig:fig4_10_10}(a) and (b) show the variation of $Q_i$ as a function of  $\langle n_{\mathrm{ph}} \rangle$ for resonators $f_{r1,s1}$ and $f_{r2,s1}$, respectively, at these temperatures.\par
At low power levels, $Q_i$ increases with temperature, consistent with thermal saturation of TLS defects, which reduces their ability to absorb energy from the resonator field. In contrast, at higher temperatures, $Q_i$ becomes a plateau at this power range, indicating that the loss becomes power-independent. This behaviour supports the interpretation that thermally activated TLS defects are no longer resonantly coupled to the microwave field, resulting in reduced sensitivity to input power.\par
The similarity in behaviour across both resonators, despite differences in absolute $Q_i$ values, points to variations in local TLS distributions or surface/interface conditions arising from fabrication. As shown in Fig.~\ref{fig:fig4_10_10}(c), $Q_c$ remains relatively constant across all powers and temperatures, confirming stable external coupling throughout the measurements. The $Q_l$ in Fig.~\ref{fig:fig4_10_10} (d) closely follows the behaviour of $Q_i$, further indicating that internal losses are the dominant limiting factor in these devices.
\section*{Discussion}
Recent studies have analysed loss mechanisms in gatemon qubits based on InAs quantum wells on InP substrates, identifying microwave losses arising from dielectric layers and defects in the quantum well and buffer layers \cite{strickland2024characterizing, casparis2018superconducting}. These losses limit the coherence times of InAs/InP-based gatemon qubits, typically to around 1 $\mu$s, which is an insufficient coherence time for practical quantum computing architectures. Therefore, developing quantum well heterostructures with low-loss substrates and buffer layers is critical for scalable superconducting qubit applications.\par In parallel, Si/SiGe heterostructures have emerged as promising candidates for semiconductor-based quantum circuits due to their compatibility with CMOS fabrication processes and potential for reduced microwave loss. In particular, the high silicon content in Si/SiGe stacks, unlike the III–V InAs-based quantum well heterostructure, may help suppress dielectric and interface-related losses, making them a more appropriate choice for low-loss integration. For instance, conventional transmon qubits fabricated on Si–Si$_{0.85}$Ge$_{0.15}$ stacks layers have demonstrated relaxation and fidelity times exceeding 100 $\mu$s, comparable to those fabricated on high-resistivity Si substrates \cite{sandberg2021investigating}. 
Relaxation times reported for voltage-tunable transmon qubits vary depending on the Josephson junction transistor structures. For instance, gatemon qubits based on Ge/Si core/shell nanowires have achieved $T_1$ times on the order of 1.27 $\mu$s \cite{zheng2024coherent}, whereas devices fabricated on Ge/Si$_{0.3}$Ge$_{0.7}$ wafers have demonstrated shorter relaxation times of approximately 80 ns \cite{sagi2024gate}. These results support that reducing the Ge concentration may help mitigate disorder-induced losses, suggesting a path toward higher-coherence, voltage-tunable devices.
In this work, the highest achievable relaxation time for the gatemon qubit using this wafer and with this design is calculated as $T = \frac{Q_l}{2 \pi f_r} \approx $ 14 ns, which indicates that further improvements are needed to achieve scalable transmon qubits with this wafer.  While this value falls short of what is needed for scalable architectures, it is comparable to recent reports on Ge/SiGe-based gatemons \cite{sagi2024gate}. Therefore, a few design improvements and optimising $Q_c$ can lead to higher coherence time in prospective voltage-tunable qubits based on Si/SiGe wafer.
On the other hand, the importance of growing Si/SiGe wafers with lower microwave losses is undeniable. CPW resonators offer a convenient testbed to evaluate and iterate on wafer growth strategies, empowering systematic research of material-induced losses. Our results emphasise that dominant loss channels are likely to come from impurity and interface roughness. As such, a careful balance between maintaining high mobility and minimising microwave loss is essential \cite{laroche2015scattering, madhavi2000low}. Additionally, reducing the thickness of the top buffer layer may mitigate the influence of scattering loss coming from the strained quantum well region due to the superconducting proximity effect, an important factor that must be considered in wafer growth.
Recent progress on isotopically enriched Si$^{28}$/Si$^{28}$Ge wafers has demonstrated low-loss quantum well heterostructures, opening a promising route toward voltage-tunable transmon qubits using CMOS-compatible materials \cite{degli2024low}. Continued efforts to understand and optimise the loss mechanisms in these wafers are essential to improve material quality and fabrication processes, ultimately improving the performance of CMOS-compatible superconducting hybrid quantum computing circuits.
\section*{Conclusion}
In this work, we developed a CMOS-compatible hybrid quantum circuit platform that combines NbN coplanar waveguide resonators with n-doped Si/Si$_{0.7}$Ge$_{0.3}$ quantum wells, achieving reproducible high internal quality factors and operational stability over more than two years at temperatures up to 1.7 K. This long-term performance highlights the reliability of our architecture for scalable quantum hardware. Through a systematic analysis of microwave loss mechanisms, we identified charge noise from the quantum well and scattering centres at substrate–vacuum interfaces as the dominant contributors, and demonstrated that targeted fabrication strategies, such as deep reactive-ion etching to recess CPW gaps, can substantially mitigate these losses. The ability to integrate these superconducting resonators with Si/SiGe wafers bridges the gap between conventional semiconductor electronics and superconducting qubits, while preserving compatibility with established semiconductor manufacturing. Furthermore, the intrinsic magnetic-field tolerance and elevated-temperature robustness of NbN make this approach well-suited for coupling to spin qubits, topological devices, and other complex quantum architectures. By enabling low-loss, manufacturable, and sustainable fabrication of superconducting–semiconducting quantum circuits, this work provides a scalable pathway towards unifying superconducting, semiconductor, and topological qubits within next-generation large-scale quantum computing systems.


\clearpage 

%

%
%
%
%
%
%


\section*{Acknowledgments}
This study made use of the University of Glasgow's James Watt Nanofabrication Centre (JWNC), and Kelvin Nanocharacterisation Centre (KNC). We thank the technical staff for their support.
\paragraph*{Funding:}
This work was supported in part by the Royal Academy of Engineering (LTRF2223-19-138), the Royal Society of Edinburgh (319941), and the Royal Society
(RGS222168), EPSRC PRF-11-I-08, and EPSRC EP/X025152/1.
\paragraph*{Author Contributions:}
P.F. designed and simulated the devices, carried out the fabrication, performed the cryogenic measurements, analysed the data, and wrote the original manuscript under the supervision of K.D. S.K. and S.P-K. contributed supporting measurements and assisted with reviewing and editing the manuscript. D.P. supplied the silicon–germanium wafer used in this study. M.W. facilitated access to the cryogenic measurement facilities. K.D. conceived the idea, supervised the project, provided research guidance, and contributed to reviewing and editing the manuscript.




\newpage


\renewcommand{\thefigure}{S\arabic{figure}}
\renewcommand{\thetable}{S\arabic{table}}
\renewcommand{\theequation}{S\arabic{equation}}
\renewcommand{\thepage}{S\arabic{page}}
\setcounter{figure}{0}
\setcounter{table}{0}
\setcounter{equation}{0}
\setcounter{page}{1} 

\newpage

\section*{Appendix}

\subsection*{Cryogenic setup for microwave spectroscopy}

\begin{figure}[t]
 \centering
 \includegraphics[width=3.0in]{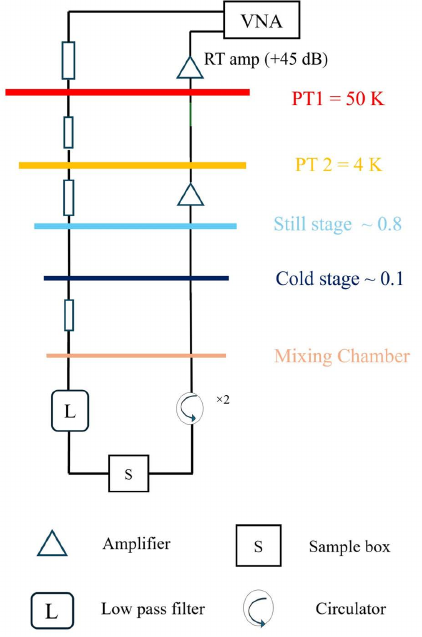}
 \caption{ Schematic illustrating the cryogenic setup employed for low-temperature measurements.}
 \label{fig:fig4_dr}
 \end{figure}

Figure~\ref{fig:fig4_dr} presents a schematic of the measurement setup, which is used throughout this work. Measurements are conducted in an Oxford Instruments Triton 200 dilution refrigerator.
The samples are glued inside an aluminium sample holder and connected to the printed circuit board with aluminium wires. Probe signals are sent from a Keysight VNA with input power from port 1, ($P_{VNA}$), attenuated, $P_{att}$, by a total of 80 dB through the stages of the DR and room temperature attenuation. Following that, the signals were delivered to the sample, and after passing through the sample, they were transmitted through an isolator. Then the signal was amplified by a low-noise amplifier mounted at the 4 K plate (+40 dB), and further amplified by a room-temperature amplifier (+45 dB).

\subsection*{Photon number calculation}
In this section, we present the model used to fit the $S_{21}$ parameter and extract the properties of the CPW resonators. The following equation gives a general model for notch-type resonators:
\begin{equation}
S^{notch}_{21} (f) = \underbrace{ a e^{i\alpha} e^{-2\pi i f\tau}}_{environment \: noise \: model} \times \underbrace{[1-\frac{(\frac{Q_{l}}{|Q_{c}|}) e^{i\phi}}{1+ 2iQ_{l}(\frac{f}{f_r}-1)}]}_{Ideal \: \:  Notch \: \:  Type \: \:  Model}
\label{Eq_notch}
\end{equation}
The prefactor in Eq.~\ref{Eq_notch} accounts for microwave background noise from the measurement setup. Specifically, $a$ represents the cable damping effect in $S_{21}$, while $\alpha$ models an additional phase rotation in $S_{21}$ due to the cable length. The electronic delay $\tau$ shifts the initial phase of the input signal $S_{21}$ from zero, caused by factors such as cable length and the finite speed of light.
The second part of Eq.~\ref{Eq_notch} describes an ideal notch-type resonator, where $f$ is the probe frequency of VNA and $\phi$ represents the impedance mismatch. In an ideal notch-type RLC circuit, $\frac{Q_l}{|Q_c|}$ can be extracted from the diameter, $d$, of the circle plot of Im($S_{21}$) versus Re($S_{21}$). Furthermore, $\phi$ indicates the phase shift of the resonator signals from the real axis due to impedance mismatch with the transmission line.\par
Based on the extracted resonator parameters, the $< n_{ph} >$ inside the CPW resonators is calculated using the following relations \cite{probst2015efficient}:
\begin{equation}
P_{in} = P_{trans} + P_{reflection} + P_{loss}
\end{equation}

\begin{equation}
P_{loss} = P_{in} (1 - |S_{21}|^2 - |S_{11}|^2)
\end{equation}

\begin{equation}
|S_{21}| = |Q_{c}|^{-2} (|Q_{c}|^2 +Q_{l}^2 - 2 Q_l |Q_{c}|)
\end{equation}

\begin{equation}
|S_{11}| = Q_l^2/|Q_c|^2
\end{equation}

\begin{equation}
< n_{ph} > = Q_i \times \frac{P_{loss}}{\hbar \omega^2}
\end{equation}
\par
Where $P_{in} = P_{VNA} + P_{att}$ is the input power inside the chip, $P_{trans}$ is the transmitted power through the feedline, $P_{loss}$ is the source of losses in the chip.
\begin{figure}[!t]
 \centering
 \hspace{-1 cm}\includegraphics[]{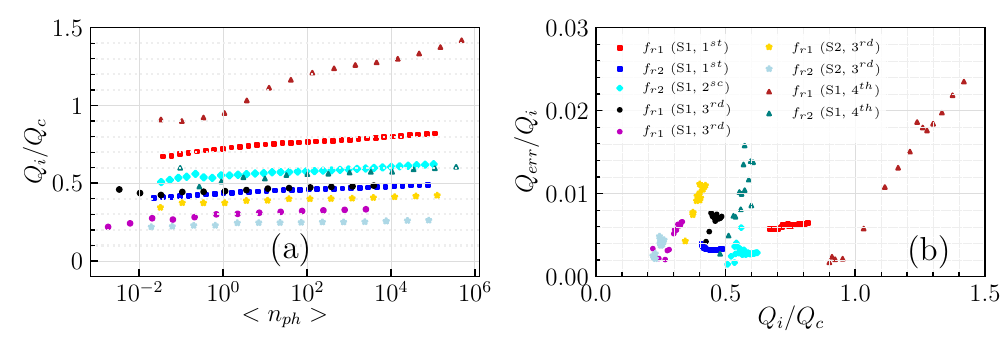}
 \caption{(a,b) Coupling coefficient versus photon number and Relative fit error ($\frac{Q_{err}}{Q_{i}}$) versus coupling coefficient at different cooldowns for both chips.}
 \label{fig:fig4_un}
 \end{figure}
\begin{figure}[!b]
  \centering
  \includegraphics[width=6.0in]{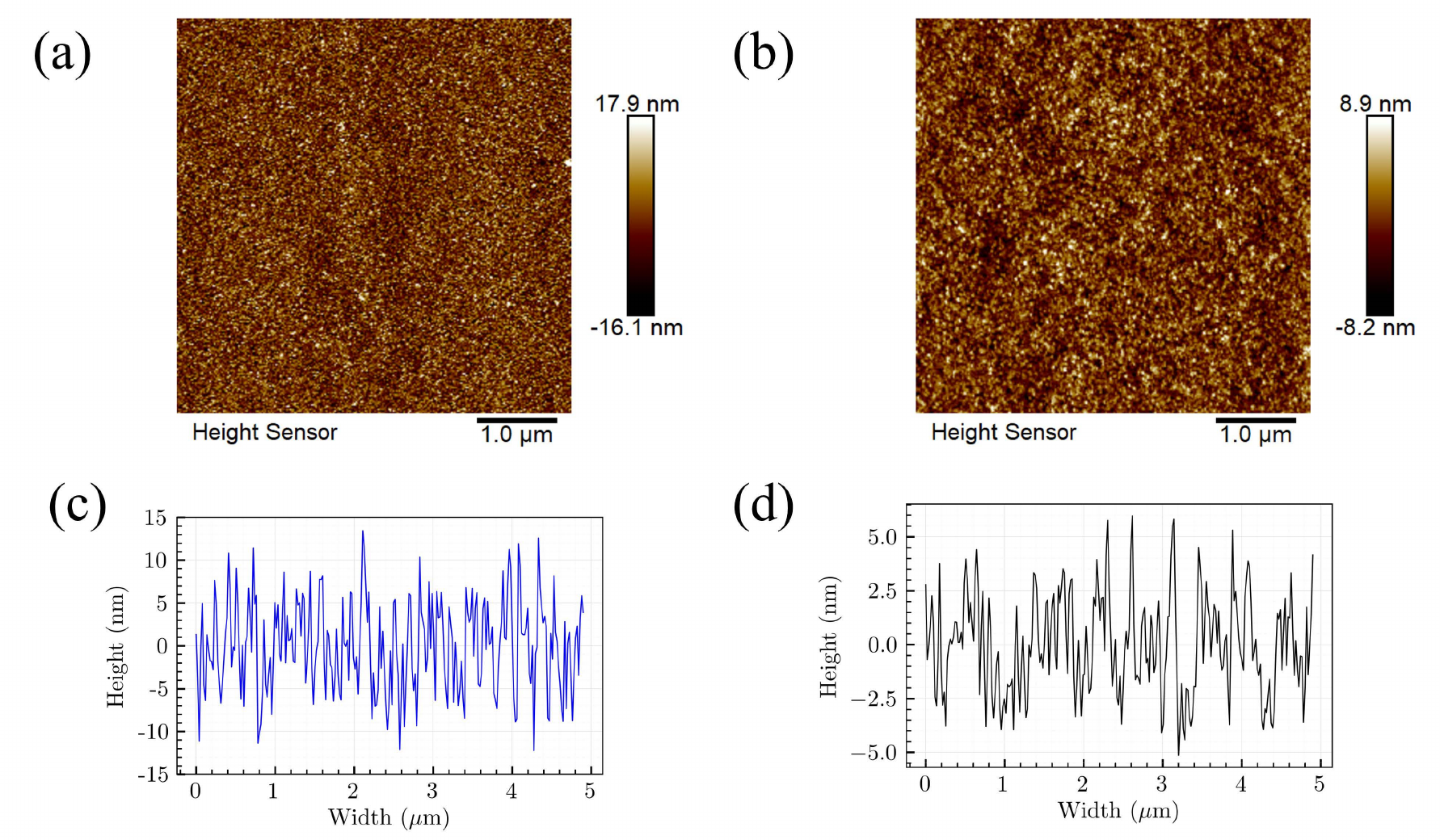}
  \caption{(a) AFM image of CPW resonators made with SiGe wafer with Roughness $\approx$ 3.9 nm. (b) AFM image of the width of CPW resonators fabricated with high resistivity Si with Roughness $\approx$ 2.17 nm. (c,d) The map shows the line profile taken in a horizontal direction.}
  \label{fig:fig4_AFM}
\end{figure}
\subsection*{Uncertainty range}
Noise sources in the CPW resonators contribute to a subtle uncertainty in the measured $Q_i$, denoted as $Q_{err}$. These noise sources may include thermal excitations, frequency noise arising from the measurement setup, and fluctuations associated with TLS. In a notch-type circuit model, the coupling coefficient $\frac{Q_i}{Q_c}$ plays an important role in determining the extent of this error \cite{baity2024circle}. When $\frac{Q_i}{Q_c}$ lies within the range of $0.1$ to $10^2$, the resulting error in $Q_i$ is minimal. Our design extracted the coupling coefficient between zero and two, shown in Fig.~\ref{fig:fig4_un} (b), resulting in a relatively small error factor in the obtained $Q_i$, as illustrated in Fig.~\ref{fig:fig4_un} (b). Figure~\ref{fig:fig4_un} (b) confirms that the CPW resonators' properties extracted from the measurement results are not considerably affected by measurement uncertainties or noise contributions.
\begin{figure}[!t]
  \centering
  \includegraphics[width=6.0in]{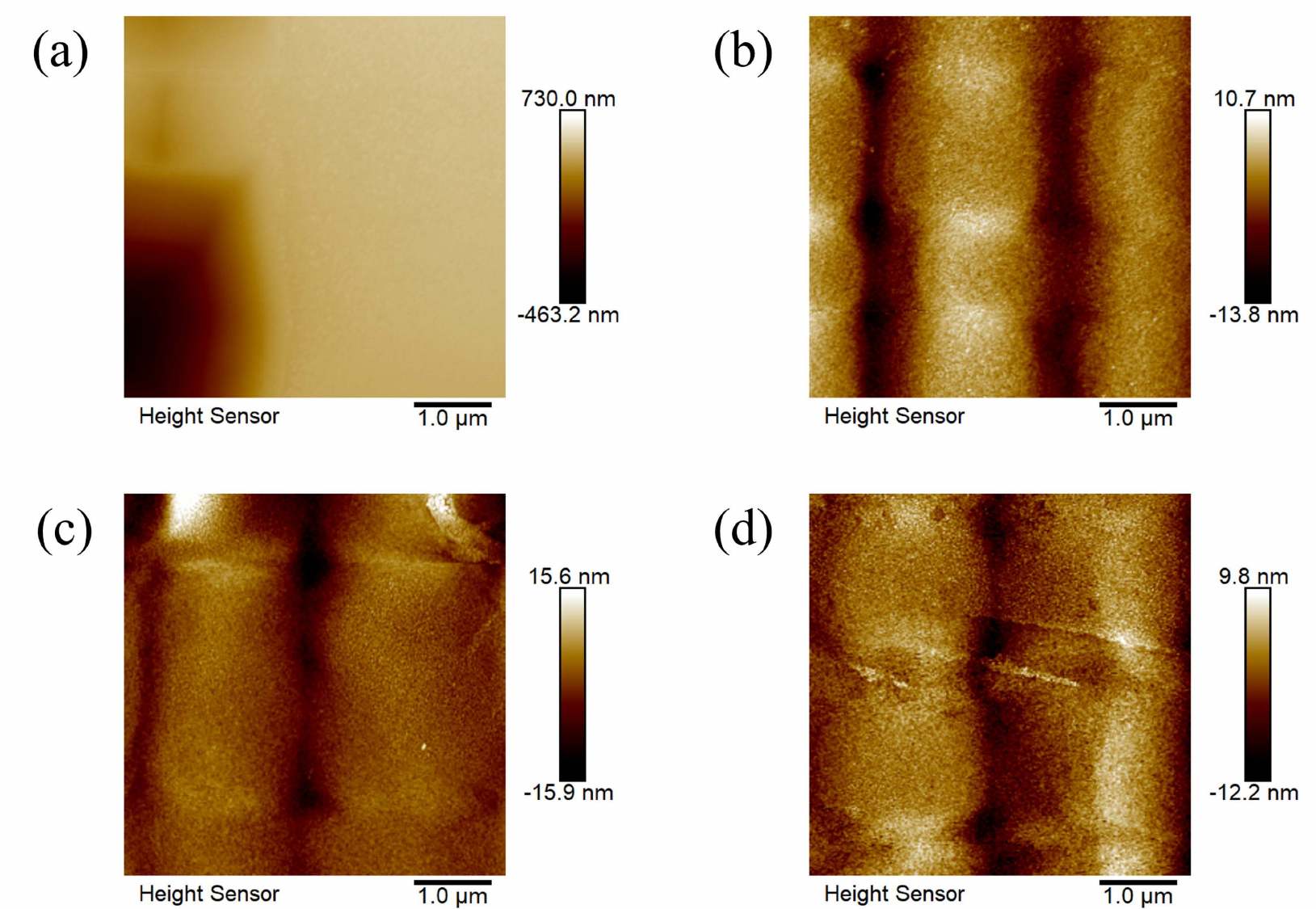}
  \caption{(a,b,c,d) AFM image of induced dislocations in the SiGe CPW resonators.}
  \label{fig:fig4_AFM2}
\end{figure}
\subsection*{AFM topography}
As mentioned, surface roughness can impact the properties of superconducting circuits, particularly $Q_i$ of CPW resonators. Figure~\ref{fig:fig4_AFM} (a) and Fig.~\ref{fig:fig4_AFM} (b) demonstrate surface roughness of CPW resonators fabricated on Si/SiGe wafer and high resistivity Si substrates \cite{foshat2025characterizing}. These figures show that the surface roughness of CPW resonators on Si/SiGe wafers is higher compared to CPW resonators fabricated on high-resistivity Si substrates. Increasing the roughness value introduces surface loss to the $Q_i$ of the CPW resonators. Comparing Fig.~\ref{fig:fig4_AFM} (c) and Fig.~\ref{fig:fig4_AFM} (d), we observe that the NbN film fabricated on high-resistivity Si exhibits greater uniformity compared to the NbN film fabricated on Si/SiGe wafers. Non-uniformity in the NbN film leads to the trapping of defects in CPW resonators and decreases $Q_i$. 
Indeed, several spots of the NbN films sputtered on Si/SiGe wafers' AFM images are illustrated in Fig.~\ref{fig:fig4_AFM2}. Figure~\ref{fig:fig4_AFM2} reveals the presence of induced dislocations within the NbN layer. These dislocations tend to add non-uniformity with thickness variation between $\approx \pm 12$ nm; however, in severe cases, this can increase to $\pm$ 500 nm or more. All these factors can contribute to increased loss in superconducting-semiconducting circuits. Therefore, it is crucial to find a way to mitigate it.




\end{document}